\documentclass[]{aa}  

\usepackage{graphicx}

\usepackage{txfonts}

\usepackage[breaklinks=true, colorlinks=true,linkcolor=blue, citecolor=blue]{hyperref}

\usepackage{layouts}
\usepackage{multirow}
\usepackage{enumitem}
\setlist{nolistsep}

\begin{document}

\title{Three-equation turbulent convection models in classical variables }

 \author{Gábor B. Kovács
          \inst{1,2,3,4}
          ,
          Róbert Szabó\inst{2,3,4}
          \and
          János Nuspl\inst{2,4}
          }

   \institute{             ELTE E\"{o}tv\"{o}s Lor\'{a}nd University, Gothard Astrophysical Observatory, Szombathely, Szent Imre h. u. 112., H-9700, Hungary
    \and
   HUN-REN Research Centre for Astronomy and Earth Sciences, Konkoly Observatory, MTA Centre of Excellence, 
              Konkoly Thege Miklós út 15-17., H-1121 Budapest, Hungary\\
              \email{kovacs.gabor@csfk.org}
            \and
             Eötvös Loránd University, Institute of Physics and Astronomy, Pázmány Péter sétány 1/a,  H-1117 Budapest, Hungary
             \and
    MTA–HUN-REN CSFK Lend\"ulet "Momentum" Stellar Pulsation Research Group, Konkoly Thege Miklós út 15-17., H-1121 Budapest, Hungary\\
             }

   \date{Received August 27, 2025; accepted December 17, 2025}

\authorrunning{G. B. Kovács, R. Szabó \& J. Nuspl}

  \abstract
   {Turbulent convection models in nonlinear radial stellar pulsation models rely on an extra equation for turbulent kinetic energy and fail to adequately explain mode-selection problems. Since multidimensional calculations are computationally expensive, it is reasonable to search for generalizations of physically grounded 1D models that approximate multidimensional results with sufficient accuracy, at least in a given parameter range.
   A natural way of progressing from one-equation models is to use additional nonlocal equations. While these types of models also exist in the literature, they have not been adopted for this type of object.}
   {We aim to adapt the three-equation turbulent convection model from Kuhfuss to radial stellar pulsation modeling.}
   {We use a Reynolds-stress one-point closure approach to derive our extensions alongside the model, while using additional models from the literature to close the anisotropy and dissipation terms.}
   {We provide five extensions to the original model. These include an enhanced dissipation correction to the mixing length, a local anisotropy model replacing eddy viscosity, a second-order correction for turbulent ion transport in the atmosphere (alongside opacity effects), and turbulent damping of entropy fluctuations and convective flux.}
   {}


   \maketitle
\nolinenumbers
\section{Introduction}

Modeling turbulent convection inside stellar cores and envelopes is one of the most important problems in astrophysics. While most studies have focused on phenomenological improvements to the mixing-length theory of \citet{mlt} \citep{kupka}, in the case of radial stellar pulsations, time-dependent convection models were necessary \citep{Houdek2015} to describe the significant effects of the narrow convective zones in the envelope of these stars. In these objects, envelope convection is confined to partial ionization zones, which also drive radial pulsations via the $\kappa$-$\gamma$ mechanism. Despite the envelope being mostly radiative, these zones provide major damping for the pulsations.

The most straightforward way of modeling stellar convection is to use 3D numerical hydrocodes. \citet{SPHERLS1,SPHERLSII} developed the multidimensional stellar pulsation code SPHERLS, which was the first to successfully model RR Lyrae stars at maximum amplitudes in 2D \citep{SPHERLSII,SPHERLSIII} and 3D \citep{SPHERLS4}, although this was achieved using relatively simple numerical methods\footnote{For a recent overview and calibration of SPHERLS, see \citet{KovacsGB2025}}. \citet{Mundprecht2013} extended the ANTARES \citep{Muthsam2010} code to model radial stellar pulsations; they compared the 2D structure to 1D treatments for a Cepheid model \citep{Mundprecht2015}. \citet[and references therein]{Vasilyev2017,Vasilyev2018} used the COBOLD5 software to model a 2D Cepheid variable and then used it as input for spectral synthesis models. Recently, \citet{Stuck2025} performed a 2D model analysis of the convective region in Cepheid models using the MUSIC code, while \citet{KovacsGB2025} performed a similar study of RR Lyrae stars using the SPHERLS code. However, both studies investigated these stars in the static limit.

All of the aforementioned studies either focus on a special case or calculate only a few stellar models due to the extreme computational cost of these simulations. Multidimensional simulations of these stars remain challenging, because adaptive grids are required to keep the partial ionization regions (which drive the pulsation) resolved as the star undergoes rapid expansion and contraction. Meanwhile, to reach maximum amplitudes, one must simulate several (thousands) pulsation cycles.
This makes improving 1D convection theories an important task \citep{kupka,KovacsGB2023}, as these models can be used in large model surveys \citep[e.g.,][]{Smolec2016} and also serve as inputs for multidimensional simulations.

Traditionally, asteroseismology has been studied primarily using linear approximations \citep{asteroseismology-book}, which have also provided important information on radial pulsators \citep{Cox-konyv}.
While linear theory can provide accurate pulsation frequencies, pulsation amplitudes enter the equations only as a normalization factor. Linear pulsation analysis can be performed in the adiabatic approximation, in which case the theory provides only frequencies. Linear nonadiabatic calculations also provide information about the growth rates of pulsation modes (i.e., whether a mode is pulsationally unstable) and result in somewhat different frequencies. Meanwhile, one can only approximately describe coupling between modes through amplitude equations. Nonlinear calculations, on the other hand, include every amplitude-dependent effect present in the system (and more accurate coupling with convection). This modifies the linear frequencies and provides additional results comparable to observations, such as light curves, amplitudes, and the actual nonlinearly selected pulsational modes. This is particularly important, because we often observe monomode pulsators where multiple modes are linearly excited. Therefore, mode selection studies investigate the nonlinear behavior of pulsation modes to determine which pulsation mode is excited, which survives if multiple modes are present, or whether they can coexist in double-mode pulsation. This, along with other nonlinear phenomena, has been central to radial pulsation modeling over the past two decades (see, e.g., \citealt{bpf-beat2002};\citealt{bpf-drrlyr2004};
\citealt{Smolec2016}). Nevertheless, both linear and nonlinear studies have shown that including turbulent convection is crucial for describing radial stellar pulsations \citep{Baker1987}.

Building on the pioneering work of \citet{Gough1965}, \citet{Unno1967}, and \citet{Castor1968}, the nonlocal models\footnote{These are the so-called one-equation models, which use a single additional nonlocal equation describing the specific turbulent kinetic energy (TKE) in the stellar envelope.} of \citet{Stellingwerf1982a}, \citet{Kuhfuss1986}, and \citet{GW1992} are used in current state-of-the-art nonlinear radial stellar pulsation hydrocodes developed in the 1990s and early 2000s \citep{Bono1994,bpf-beat2002,lengyel,Paxton2019}. For a detailed summary, see \citet{KovacsGB2023}.

Meanwhile, these models have their own limitations: they have many free parameters; nonlinear phenomena such as mode selection and double-mode pulsations
are model-dependent \citep{bpf-drrlyr2004,lengyel2,KovacsGB2023,KovacsGB2024}; light and radial velocity curves show discrepancies with observations \citep{Marconi2017rev,KovacsGB2023} and the models produce unphysical stratification in the stellar envelope \citep{Canuto1998,Kupka2022,KovacsGB2023}.
  
Many of these problems stem from the so-called downgradient approximation of the convective flux and the resulting nonphysical stratification of turbulent kinetic energy (TKE) (\citealt{Canuto1998}; \citealt{Kupka2022}; \citealt{KovacsGB2023}). 
In the downgradient approximation, the convective flux is considered proportional to the super-adiabatic gradient. There is a further assumption that in stable regions (i.e., where the gradient is negative), the gradient is taken as zero \citep{Stellingwerf1982a}. This approximation also affects the TKE profile, which in turn influences mode selection through the turbulent pressure and viscous stress terms. This effect determined whether simulations reproduce or fail to exhibit double-mode pulsation \citep{lengyel,lengyel2,KovacsGB2023}. Neglecting negative convective flux leads to insufficient turbulent damping, causing overshooting into the deep stellar interiors. This provides additional damping that can stabilize double-mode pulsations \citep{lengyel2,KovacsGB2023}. Meanwhile, maintaining a negative flux prevents TKE leakage to deeper regions but leads to unstable double-mode pulsations, where one mode is eventually damped despite a ``physically more correct'' picture. This occurs because negative flux terms cause strong turbulent damping in the overshooting region, where convective elements behave as if colliding with a rigid wall \citep{Kupka2022,KovacsGB2023}. Both pictures are nonphysical, illustrating the limitations of one-equation models. While the aforementioned problems depend primarily on the coupling between convection and radial pulsation, other observed features -- such as the Blazhko effect \citep{Blazhko1907} and additional modes \citep[the so-called $f_x$ modes; see, e.g.,][]{Szabo2015} -- can be explained by resonances with non-radial modes (see, e.g., \citealt{VanHoolst1998}; \citealt{Dziembowski2004}; \citealt{Dziembowski2016}).
Although such resonances require multidimensional treatment, improving the 1D description cannot be ruled out as a means to explain these features. 

The analysis of multidimensional models by \citet{Mundprecht2013,Mundprecht2015} and \citet{KovacsGB2025} reveals that a natural progression is to incorporate nonlocal time-dependent differential equations describing the convective flux and temperature fluctuations. Developed by \citet{Xiong1980b,Xiong2021,Xiong1998a} and \citet{Kuhfuss1987}, these so-called three-equation models were incorporated into stellar evolution codes \citep[see][]{Flaskamp2002,Flaskamp2003,Weiss2008} and used in linear non-radial pulsation theory \citep[][and references therein]{Xiong1998b,Xiong2000,Xiong2016,Xiong2018}. However, further adjustments to these models are necessary,
since convection in classical pulsators substantially differ from standard astrophysical convection: 
\begin{itemize}
    \item Highly dynamical regime. The convection timescale is comparable with the cycle of the pulsation, making a time-dependent description important, while the large structural changes during pulsation create more complex coupling between pulsation and convection.
    \item Anisotropy effects. Convective eddies convert kinetic energy into TKE through shear. This eddy viscosity is crucial in damping convection \citep{Gonczi1981,Stellingwerf1982a}, and it arises from the anisotropy of convective motions \citep{GW1992}. Consequently, modeling anisotropies is also important \citep{Mundprecht2015}.
    \item Sharp ionization regions, especially in the case of RR Lyraes. Most of these stars exhibit a sharp HI ionization region.
    Small perturbations in ionization levels have a drastic effect on specific heat and opacity, which must be taken into account \citep{KovacsGB2025}.
\end{itemize}

We introduce an extended version of the Kuhfuss three-equation model in this paper \citep{Kuhfuss1987}\footnote{It is available through the Universit\"{a}tsbibliothek
Technische Universit\"{a}t M\"{u}nchen at \url{https://mediatum.ub.tum.de/doc/1718437/1718437.pdf}}, which provides a basis for improving 1D nonlinear stellar pulsation codes.

\section{Extended Kuhfuss three-equation model}
\label{sec:model}
We provide the basic equations of the extended \cite{Kuhfuss1987}
 \ model below, while a derivation consistent with our notation is given in Appendix \ref{ap:derive}. The equations are the following:
\begin{align}
    \partial_t \rho + \frac{1}{r^{2}}\partial_r (r^2\rho v) &= 0,\label{eq:model_continuity}\\
    {\rm d}_t v& = -\frac{1}{\rho}\partial_r(p+p_t) - U_\nu-g ,\\
    {\rm d}_t e + p\frac{1}{r^2}\partial_r (r^2v) &= -\frac{1}{\rho r^2}\partial_r\left[r^2\left(F_{\rm r}+F_{\rm c}\right)\right]-\frac{2T\rm \Phi}{c_p\tau_\kappa}-S+\varepsilon,\label{eq:E}\\
    {\rm d}_t\tilde{\omega} + p_t\frac{1}{r^2} \partial_r(r^2v)&=-\frac{1}{\rho r^2}\partial_r\left(r^2\mathcal{F}_{\tilde{\omega}}\right) + E_\nu + S - \varepsilon ,  \\
    {\rm d}_t{\rm \Phi} &=-\frac{1}{\rho r^2}\partial_r\left(r^2\mathcal F_{\rm \Phi}\right)+\mathcal P_{\rm \Phi} - \epsilon_{\Phi}- \frac{2\Phi}{\tau_\kappa},\\
    {\rm d}_t{\rm \Pi} &= -\frac{1}{\rho r^2} \partial_r \left(r^2 \mathcal{F}_{\rm \Pi}\right) + \mathcal{P}_\Pi + S_\Pi-\epsilon_\Pi -\frac{\Pi}{\tau_\kappa}\label{eq:model_pi}.
\end{align}
Here, ${\rm d}_t = \partial_t + v\partial_r$ denotes the Stokes derivative, and $\partial_r$ is the spatial derivative relative to the $r$, radius coordinate\footnote{The system of equations can be converted onto a Lagrangian grid with mass coordinates using the mass coordinate defined as ${\rm d}m =4\pi r^2\rho {\rm d} r$ \citep{Kippenhahn-book}.}. The thermodynamic quantities follow standard notation: $p$ is pressure, $\rho$ is the density, $T$ is the temperature, $e$ is specific internal energy, and $c_p$ is specific heat at constant pressure. The variable $v$ describes the velocity of a given spherical shell, while $g=Gm/r^2$ denotes the gravitational acceleration, where $m$ is the mass inside a sphere of radius $r$ within the star. $F_r$ is the radiative energy flux in the diffusion approximation:
\begin{displaymath}
    F_r = K_r \partial_rT = \frac{16\sigma_{\rm B}T^3}{3\rho \kappa}\partial_r T,
\end{displaymath}
where $\kappa(T,\rho)$ is the Rosseland-mean opacity and $\sigma_{\rm B}$ is the Stefan-Boltzman constant. The remaining quantities relate to turbulent convection and its coupling to the dynamics. The specific TKE is denoted by $\tilde{\omega}$, the entropy fluctuation is denoted by $\rm \Phi$, and $\rm \Pi$ denotes the velocity entropy covariance.

The convective flux, $F_c$, is represented by a second-order description:
\begin{equation}
    F_c = T \rho\Pi + \alpha_c\frac{c_{pT} T}{c_p}\mathcal{F}_{\rm \Phi}, \label{eq:Fc}
\end{equation}
where $c_{pT}=(\partial\ln c_p/\partial \ln T)_p$ and $\alpha_c$ is a dimensionless factor of order of unity.

The term $p_t= (2/3)\rho\tilde{\omega}$ denotes turbulent pressure, while terms $E_\nu$ and $U_\nu$ denote the turbulent viscous energy and momentum transfer rates \citep{Wuchterl1998}:
\begin{align}
    U_\nu &= \frac{1}{\rho}\partial_r\left[\left(\xi-\frac{1}{3}\right) 2\tilde{\omega}\rho\right] + \left(\xi-\frac{1}{3}\right)\frac{6\tilde\omega}{r}, \\
    E_\nu &= 2\tilde\omega \left[\left(\xi-\frac{1}{3}\right)\partial_rv+ 2\left(\frac{1}{3}-\xi\right)\frac{v}{r}\right].
\end{align}
Here, $\xi$ denotes the flow anisotropy (see Appendix \ref{ap:derive}). In eqs. (\ref{eq:E})-(\ref{eq:model_pi}), we refer to buoyancy-driven terms as source terms denoted by $S$, while shear-related terms are referred to as production terms \citep{Pope-konyv} denoted by $\mathcal{P}$:
\begin{align}
    S &= - \frac{\delta}{\rho c_p}{\rm \Pi} \partial_r p,
    &S_{\rm \Pi} &= - \frac{\delta}{\rho c_p}\rm{\Phi} \partial_r p,\\
    \mathcal{P}_\Phi &= - {\rm \Pi} \partial_r s,
    &\mathcal{P}_\Pi &=-2\xi\tilde{\omega}\partial_r s,
\end{align}
where $\partial_r s$ is the specific entropy.

The dissipation terms are modeled locally through the dissipation and radiative timescales. First, the dissipation of TKE ($\varepsilon$) is modeled using the convective timescale. This is constructed from a characteristic length (i.e., mixing length) and characteristic velocity (approximated as $\sqrt{2\omega}$). This yields $\tau_{\rm conv} = \Lambda/(2\tilde\omega)^{-1/2}$, $\varepsilon = \tilde\omega/\tau_{\rm conv}$, and
\begin{equation}
    \varepsilon = \alpha_d \frac{\tilde\omega^{3/2}}{\Lambda}, 
\end{equation}
where the factor $\sqrt2$ is incorporated into the dimensionless constant $\alpha_d$ of order unity. The dissipation of the entropy fluctuations ($\epsilon_\Phi$) and velocity-entropy covariance ($\epsilon_\Pi$) has two components: a radiative ($\propto \tau_{\rm r}^{-1}$) and a viscous part ($\propto \Lambda^{-1}$):
\begin{align}
    \epsilon_\Phi &= \frac{2\Phi}{\tau_r} + \alpha_{d,\Phi}\frac{2\Phi\tilde\omega^{1/2}}{\Lambda},
    &\epsilon_\Pi &= \frac{\Pi}{\tau_r} + \alpha_{d,\Pi} \frac{\Pi \tilde\omega^{1/2}}{\Lambda}. \label{eq:dissipation}
\end{align}

The radiation timescale can be determined from the Péclet number, since $Pe = \Lambda \sqrt{2\tilde\omega}\rho c_p/K_r = \tau_r/\tau_{\rm conv}$. Thus,
\begin{equation}
    \tau_r = Pe \times \tau_{\rm conv} = \alpha_r^{-1} \frac{\rho^2\Lambda^2c_p\kappa}{\sigma_{\rm B} T^3},
\end{equation}
where $\alpha_r = 3/16$, is a dimensionless constant\footnote{Here we use the reciprocal of $\alpha_r$ to simplify the parameterization of the equations.}.

The final timescale introduced is the opacity-induced dissipation-production timescale, which provides additional dissipation or production of entropy fluctuations. It is defined by $\tau_\kappa = \tau_r/(3-\kappa_T)$, where $\kappa_T=(\partial \ln \kappa/\partial \ln T )_p$.

Finally, the third-order moments (TOMs) describe the turbulent flux of turbulent quantities and are approximated by the down-gradient approximation\footnote{Local equations for the TOMs can also be derived by localizing the full nonlocal equations of \citet{Canuto1992}. However, this method introduces new variables that complicate numerical solutions. Meanwhile, to determine which terms have real effects on radial stellar pulsations, it is better to proceed one step at a time.} \citep{Castor1968,Xiong1980b}:
\begin{align}
    \mathcal{F}_{\tilde\omega} &= -\alpha_{\tilde\omega} \mu_t \partial_r\tilde\omega, &\mathcal{F}_{\rm \Phi} &= -\alpha_{\rm \Phi} \mu_t \partial_r{\rm \Phi} ,&\mathcal{F}_{\rm \Pi} &= -\alpha_{\rm \Pi} \mu_t \partial_r{\rm \Pi},
\end{align}
where $\mu_t=\Lambda \rho\sqrt{2\xi\tilde\omega}$ is the turbulent viscosity. 
\section{Modifications}

\subsection{Enhanced dissipation}

The local model of viscous dissipation, $\varepsilon$, is known to be insufficient in the dynamically stable regions \citep{Canuto1993} when using the standard prescription for the mixing length $\Lambda=\Lambda_0=\alpha_\Lambda H_p$, where $H_p=pr^2/(\rho{\rm G}m)$ is the pressure scale height and $\alpha_\Lambda$ is a dimensionless parameter. This problem arises because gravity waves provide extra damping  in the stable regions. \citet{Kupka2022} derived a correction to the mixing length, which we adopt here using the formula introduced of \citet{Ahlborn2022}:
\begin{multline}
    \Lambda(r) = \left\lbrace\begin{matrix}
    \displaystyle - \frac{r+\Lambda_0}{2\tau^\star}+\sqrt{\left[\frac{r+\Lambda_0}{2\tau^\star}\right]^2+ \frac{\Lambda_0 r}{\tau^\star}} \quad  &\text{if } (\nabla-\nabla_{\rm ad}) <0,\\ \\
    \displaystyle  \frac{\alpha_\beta r}{\Lambda_0+\alpha_\beta r} \Lambda_0&\text{if } (\nabla-\nabla_{\rm ad}) \ge 0, 
    \end{matrix} \right.
\end{multline}
where
\begin{equation}
    \tau^\star =  \alpha_\tau \tilde{\omega}^{-1/2}\Lambda_0  g \sqrt{\rho p^{-1}\left(\nabla_{\rm ad}-\nabla\right)}
\end{equation}
is the ratio of the dissipative to buoyancy timescales in the stable regions, and $\alpha_\beta$ and $\alpha_\tau$ are dimensionless parameters. Here, $\nabla-\nabla_{\rm ad} = H_p T^{-1} \partial_r s$ is the dimensionless super-adiabatic gradient, with $\nabla =(\partial \ln T /\partial\ln p)$ and $\nabla _{\rm ad}=(\partial \ln T /\partial\ln p)_s$. 

\subsection{Local anisotropy equation}

We follow the traditional splitting of the Reynolds-tensor into isotropic turbulent pressure and anisotropic turbulent viscous stress tensor. The latter is usually modeled using the eddy-viscosity hypothesis \citep{Pope-konyv}, which yields the anisotropy parameter
\begin{equation}
    \xi \approx \frac{1}{3}-\frac{\alpha_\nu\Lambda}{2\tilde{\omega}^{1/2}}\partial_r v. \label{eq:xi_simple}
\end{equation}
\citet{Kuhfuss1987} applied the above formulation only to the turbulent viscous stress tensor, setting $\xi=1/3$ elsewhere\footnote{Using Eq. (\ref{eq:xi_simple}) recovers the original definitions of $U_\nu$ and $E_\nu$ from \citet{Wuchterl1998}.}. Recently \cite{KovacsGB2023,KovacsGB2024} showed that the $\alpha_\nu$ parameter correlates with $\alpha_d$ and significantly affects mode selection \citep{KovacsGB2024} and nonlinear behavior \citep{Kollath2011}. Therefore, instead of Eq. (\ref{eq:xi_simple}), we propose a local prescription derived from the nonlocal model of \cite{Canuto1998b} using the method given in \citet{Kupka2022}, but retaining the shear terms (see appendix \ref{ap:anisotropy}):
\begin{multline}
\xi = \frac{1}{3} + \frac{\tau_\xi}{3\tilde\omega}\left[\mathcal{D}_\omega-\left(2\beta_5+1\right)S+\varepsilon \right]- \frac{2\tau_\xi}{9}\left[\left(\alpha_1-\frac{1}{5}\right)\partial_rv-2\frac{v}{r}\right],
\end{multline}
equivalent to \citeauthor{Kupka2022}'s (\citeyear{Kupka2022}) solution in the static case. Here, $\mathcal{D}_\omega =\sqrt{2/3} \alpha_t\rho^{-1}r^{-2}   \partial_r (r^2\Lambda \rho \tilde{\omega}^{1/2} \partial_r \tilde{\omega})$ describes diffusion of $\xi$ through isotropic TKE flux, $\tau_\xi^{-1}=(2/3)\left[\alpha_1(v/r)+(2\alpha_1-1)\partial_r v\right] - 5\varepsilon/(2\tilde{\omega})$, while $\alpha_1=1.08$ and $\beta_5=0.7$ are constant parameters given by \citet{Canuto1994}.

\subsection{Ionization energy transport}

The second term in Eq. (\ref{eq:Fc}) provides a second-order correction representing the ionization energy flux carried by ions transported via convection.In fact, this reflects a change in chemical potential, even though ionization effects are usually absorbed into $c_p$ and $\kappa$ \citep{Kippenhahn-book}. It becomes a significant component of the total enthalpy flux due to the sharp stratification present in RR Lyrae stars \citep{KovacsGB2025}. Since its primary effect is on specific heat, we parameterize it through temperature fluctuations (see Appendix \ref{ap:convection}).

\subsection{Opacity effects}

Sharp stratification in the ionization zones also results in strong opacity bumps; therefore, opacity fluctuations cannot be neglected. Thus, we must track changes in radiative conductivity $K_r$, approximated by a first-order Taylor series of $K_r(T,\kappa)$, and $\kappa(T)$ \citep{Xiong1980b}. While the opacity bumps enhance radiative losses, they also generate further fluctuations analogous to the $\kappa$-mechanism. This is handled through the opacity-induced dissipation-production timescale, $\tau_\kappa$.

\subsection{Viscous dissipation}

The sometimes short dynamical timescales alongside enhanced opacity (i.e., less efficient cooling) make the otherwise insignificant viscous damping of the entropy fluctuations and flux relevant. Assuming scale similarity with kinetic energy damping and the convective timescale $\tau_{\rm conv}=\tilde{\omega}/\varepsilon$, the second terms in (\ref{eq:dissipation}) follow.

\section{Discussion}

The model proposed here reduces to the standard three-equation model of \citet{Kuhfuss1987} when using (\ref{eq:xi_simple}) in $E_\nu$ and $U_ \nu$, setting $\xi=1/3$ elsewhere, neglecting $\tau_\kappa$ terms, using $\Lambda=\alpha_\Lambda H_p$, and also setting the parameters $\alpha_c$,$\alpha_{\rm d,\Pi}$,$\alpha_{\rm d,\Phi}$ to zero. Some of our extensions can also be incorporated into one-equation models, such as the local anisotropy formulation.

Most of the free dimensionless parameters can be adopted from the existing literature. Assuming isotropy in the dissipation regime, we can adopt the dissipation parameter of \citet{Kuhfuss1987} calibrated to the mixing-length theory: $\alpha_d=8/3\sqrt{2/3}\approx2.17$.

The diffusion parameters can be obtained from those of \citet{Kuhfuss1987} by dividing them by $\sqrt{2\xi}=\sqrt{2/3}$. Therefore, we obtain $\alpha_\Pi =6$ and $\alpha_\Phi = 4$. In the case of $\alpha_t$, we can start from the local condition of the one-equation model $\sqrt{3\alpha_t/(2\alpha_d)} \le (3/8)\sqrt{2}$, which gives $\alpha_t \le 4/8$.

Regarding the remaining dissipation terms, Kuhfuss sets the viscous dissipation timescale of $\Pi$ and $\Phi$ to $\varepsilon/(c_p T)$ and then neglects it in the local limit \citep{Kuhfuss1987}. This is equivalent to scaling $\alpha_{d,\Pi}\approx\alpha_{d,\Phi}$ parameters with $\alpha_d$ such that  $\alpha_{\rm d,\Pi}/\alpha_d \sim\tilde{\omega}/(c_p T)$, which is between $10^{-1}$ and $10^{-2}$ in the ionization region. Thus, we adopt $\alpha_{\rm d,\Phi} \alpha_{\rm d,\Pi} = 0.0217 \approx\alpha_d/100$.

The mixing-length parameters, $\alpha_\Lambda$ and $\alpha_\beta$, can follow the standard values used in existing 1D codes \citep{Kupka2022,KovacsGB2024}: $\alpha_\Lambda =1.5$ and $\alpha_\beta = 1$. The value $\alpha_\tau = 0.2$ is given by \citet{Kupka2022}. The authors also give $\alpha_\tau = 5/(32\alpha_d)$. These two values match only when $\alpha_d =0.8$, derived from Kolmogorov-scaling \citep{Canuto1998}.

We chose $\alpha_r=3/16$, consistent with simple Péclet number dependence for the radiative timescale, rather than the mixing-length calibrated value of \citep{Kuhfuss1987} (equivalent to $\alpha_r = 48$  in our notation).

Lastly, $\alpha_c=1$ effectively captures ionization energy transport effects (see Appendix C), whereas \citet{Montgomery2004} used $\alpha_c=0.5$ from second-order Taylor series enthalpy fluctuations.

Although our modifications improve the original \citet{Kuhfuss1987} model for nonlinear pulsation codes, we retain a mixing length approach and a local description of turbulent dissipation and anisotropy. A more complete model \citep[e.g.,][]{Canuto1998b} is not recommended at this stage, because a gradual improvement in the treatment help identify the key phenomena governing nonlinear pulsation. Similar to its predecessor, our model is based on the assumption of local thermodynamic equilibrium and neglects radiative transfer in optically thin regions. As these are essential for accurately modeling light curves, the next step should implement them into the models\footnote{For one-equation models, this procedure was performed by \citet{GW1992} as well as \citet{Dorfi1991}.}.

In conclusion, implementing this model into a new hydrocode should be done gradually. Successive modifications may likely have diminishing effects: the first two have the largest effect, followed by the convection correction and opacity effects. The effect of the additional dissipation terms is likely to have an effect only in mode selection and nonlinear phenomena. We also note that the chosen parameters given here also require observational calibration.

\section{Conclusions}

In this paper, we presented our extended version of the Kuhfuss three-equation turbulent convection model \citep{Kuhfuss1987}, tailored for use in the field of nonlinear radial stellar pulsations.
Our main modifications include:
\begin{enumerate}
    \item the adoption of the enhanced turbulent dissipation from \citet{Kupka2022},
    \item  the derivation of a local non-static anisotropy model from \citet{Canuto1998b},
    \item a second-order correction to the convective flux,
    \item a first-order correction accounting for opacity fluctuations. This and the flux correction represent effects of sharp ionization regions and opacity sensitivity \citep{KovacsGB2025}.
    \item the retention of turbulent dissipation for entropy fluctuations and flux, expected to have additional stabilizing effect on nonlinear behavior.
\end{enumerate}

Numerical comparison with state-of-the-art one-equation hydrocodes \citep[as analyzed in ][]{KovacsGB2023,KovacsGB2024} and a detailed analysis of the model approximations, based on multidimensional results, will be presented in upcoming papers.
\begin{acknowledgements}

We are grateful to the anonymous referee for the careful and thorough reading of our manuscript, and whose suggestions improved the quality of this paper.

The research was supported by the EK\"OP-24 University Excellence Scholarship Program of the Ministry for Culture and Innovation from the source of the National Research, Development and Innovation Fund, through the EK\"OP-24-4-I-ELTE-363 grant.

This project has been supported by the `SeismoLab' KKP-137523 \'Elvonal grant, Lendület program of the Hungarian Academy of Sciences project No. LP2025-14/2025, OTKA projects K-129249, SNN-147362, K-147131 and NN-129075, as well as the NKFIH excellence grant TKP2021-NKTA-64.  This research was also supported by the International Space Science Institute (ISSI) in Bern/Beijing through ISSI/ISSI-BJ International Team project ID \#24-603 - “EXPANDING Universe” (EXploiting Precision AstroNomical Distance INdicators in the Gaia Universe).

On behalf of Project 'Hydrodynamical modeling of classical pulsating variables with SPHERLS' we are grateful for the usage of HUN-REN Cloud \citep[see][\url{https://science-cloud.hu/}]{H_der_2022} which helped us achieve the results published in this paper. 

\end{acknowledgements}

\bibliographystyle{aa} 
\bibliography{bibgraph}
\begin{appendix}
\section{Deriving the model}
\label{ap:derive}
We start from the Navier-Stokes equations, of the continuity, momentum, and energy conservation in index notation. As we focus on envelope convection in a pulsating envelope, we neglect any energy production from nucleosynthesis and rotation, while we keep all time derivatives. Thus, we have:
\begin{align}
    \partial_t \rho + \partial_k(v_k\rho) &= 0,\label{eq:kont}\\
    {\rm D}_tv_j &= -\frac{1}{\rho}\,\partial_j p + \frac{1}{\rho}\,\partial_k \sigma_{kj}+ g\delta_{j3},\label{eq:mom}\\
    {\rm D}_t\left(e + \frac{1}{2}v_j^2\right)& = -\frac{1}{\rho}\,\partial_j\bigl(p\,v_j + F_j - \sigma_{jk}\,v_k\bigr)+ g_j\,v_j.\label{eq:energy}
\end{align}
Here, $\rho$ denotes the density, $v$ denotes the velocity, $p$ is the thermodynamic pressure, $\sigma$ is the viscous stress tensor, $g=-{\rm G }m/r^2$ is the gravitational acceleration, where $m$ is the mass inside the spherical shell with radius $r$. $F$ denotes radiative flux and $e$ is the internal energy. The symbol ${\rm D}_t = \partial_t + \overline v_k \partial_k$. Additionally, we define the heat sources as the radiative flux and viscous heat production, that is, the specific entropy ($s$) production is
\begin{equation}
    T{\rm D_t}s = -\frac{1}{\rho}\partial_i F_i + \sigma_{ij}\partial_i v_j. \label{eq:heat}
\end{equation}

We define Reynolds-average as the horizontal average of a quantity, namely
\[\overline{q} \coloneqq \frac{1}{4\pi}\int_0^{2\pi}\int_0^{\pi} q(r,\theta,\phi) \sin \theta{\rm d}\theta {\rm d} \phi.\]
This operation is commutative with (Eulerian) spatial and time derivatives. Using this operation, we split quantities into mean and fluctuation parts (denoted as $(\cdot)'$), for example:
\begin{displaymath}
    v = \overline{v}+v',\quad p=\overline p + p',\quad\dots
\end{displaymath}
The operator has the following identities:
\begin{displaymath}
    \overline{ab} = \overline a \overline b + \overline{a'b'},\quad \overline{a'} =0,\quad \overline{\overline a} = \overline a .
\end{displaymath}

We denote the Stokes derivative related to the average field as \citep{Kuhfuss1986}
\begin{equation}
    {\rm d}_t q = \partial_t q + \overline v_k \partial_k q = {\rm D}_t q - v'_k\partial_k q.
\end{equation}

Now we derive equations for the mean and fluctuating quantities. First, we neglect $\rho'$ except buoyancy, then write Eq. (\ref{eq:kont}) as
\begin{align}
    \partial \overline \rho +\partial_k (\overline \rho\, \overline v_k) &= 0,\label{eq:mean_kont}\\
    \partial_k (\overline \rho v'_k) &= 0. \label{eq:anel}
\end{align}
Here, (\ref{eq:anel}) is called anelastic approximation.

Taking the average of Eq. (\ref{eq:mom}), we get the momentum equation of the pulsation (i.e., large-scale motion)
\begin{equation}
    d_t \overline{v}_j + \overline{v'_k\partial_kv_j'} = - \frac{1}{\overline \rho} \partial_j \overline p +\frac{1}{\overline \rho}\partial_k\overline{\sigma}_{kj}  + g\delta_{j3} \label{eq:mean-mom1}.
\end{equation}
Here, the second term can be converted into a tensor divergence, using Eq. (\ref{eq:anel}), one can show that $\overline{v'_k\partial_kv_j'} = (1/\overline\rho)\partial_k\overline{\rho v_k'v_j'}\equiv -(1/\overline\rho)\partial\tilde{\sigma}_{kj}$, where we define the Reynolds tensor as \citep{Kuhfuss1986} 
\begin{equation}
    \tilde\sigma_{ij}= -\overline{\rho v'_i v_j'}. \label{eq:Reynolds}
\end{equation}
After rearranging, Eq. (\ref{eq:mean-mom1}).
\begin{equation}
    d_t \overline{v}_j = - \frac{1}{\overline \rho} \partial_j \overline p +\frac{1}{\overline \rho}\partial_k\left(\overline{\sigma}_{kj} + \tilde{\sigma}_{kj}\right)  + g\delta_{j3} \label{eq:mean-mom2}.
\end{equation}

We derive the momentum equation of the fluctuating field by subtracting Eq. (\ref{eq:mean-mom1}) from Eq. (\ref{eq:mom}), thus
\begin{equation}
    d_t v'_j + v_k'\partial_k \overline{v}_j + (v_k'\partial_kv_j')' = - \left(\frac{1}{\rho}\partial_j p\right)' + \frac{1}{\overline \rho}\partial_k\sigma'_{kj}. \label{eq:fluc_mom}
\end{equation}
Here, the first term of the right-hand side can be expanded as
\[-\frac{\rho'}{\overline\rho^2}\partial_j\overline p + \frac{1}{\overline \rho} \partial_j p'\],
where the second term is usually neglected in the small Mach-number approximation \citep{Gough1977}. On the other hand, \citet{Canuto1992} gives an explanation of the importance of pressure fluctuations, while \citet{Viallet2013} showed that pressure transport can be a significant contribution to envelope convection.

Next, we derive the mean total energy equation similarly to \citet{Kuhfuss1986}. Applying averaging, the left-hand side of Eq. (\ref{eq:energy}) becomes \citep{Kuhfuss1986,Kuhfuss1987}:
\begin{multline*}\overline{{\rm D}_t\left((e + \frac{1}{2}v_j^2\right)} = {\rm d_t}\left( \overline e + \frac{1}{2} (\overline v_j)^2 + \frac{1}{2}\overline{v_j'v_j'}\right) \\+ \overline{v'_k\partial_k e'}-\frac{1}{\overline \rho} \left(\tilde{\sigma}_{kj}\partial_k\overline v_j+\overline v_j\partial_k\tilde{\sigma}_{kj}\right)+\frac{1}{2}\overline{v_k'\partial_kv_j'v_j'}
\end{multline*}
Here, the Reynolds stress-related terms are derived by using Eq. (\ref{eq:anel}) and the definition (\ref{eq:Reynolds}). The term $(1/2)\overline{v_j'v_j'} =\tilde{\omega} \equiv (1/\overline\rho) \tilde{\sigma}_{jj}$ is the kinetic energy stored in the turbulence, and proved to be crucial in modeling radial stellar pulsations \citep{Stellingwerf1982b}. Therefore, it's necessary to close it through an exact equation. The third order moment (TOM) $(1/2)\overline{v_k'\partial_kv_j'v_j'} = (1/2\overline{\rho})\partial_k \left(\overline\rho\,\overline{v_k' v_j'v_j'} \right)$, where Eq. (\ref{eq:anel}) was used, and it describes the turbulent flux of the TKE. 

The RHS of the averaged (\ref{eq:energy}) is
\begin{multline*}
    -\frac{1}{\overline \rho} \partial_i\left(\overline{p}\,\overline{v}_j+\overline{p'v_j'} + \overline F_j - \overline \sigma_{jk}\overline{v}_k - \overline{\sigma_{jk}'v_k'}\right)+ g_j \delta_{j3} \overline{v}_j.
\end{multline*}

Multiplying (\ref{eq:mean-mom2}) with $\overline v_j$, one can eliminate the kinetic energy of the averaged field. The pressure fluctuation term can be put together on the right-hand side after rearranging the $\overline{v_k'\partial_ke'}$ term using Eq. (\ref{eq:anel}), because $ \overline \rho h' = \overline \rho e' + p'$, where $h$ is the specific enthalpy. Using this, the average of the total energy becomes
\begin{multline}
    {\rm d}_t\left( \overline e + \tilde{\omega}\right) = - \frac{1}{\overline \rho} \overline{p}\partial_j\overline{v}_j\\
    -\frac{1}{\rho}\partial_j\left(F_j + \overline{\rho}\,\overline{h'v_j'}+\frac{1}{2}\overline{\rho}\,\overline{v_j'v_k'v_k'}-\overline{\sigma'_{jk}v_k'}\right) + \frac{1}{\overline\rho}\left( \overline{\sigma}_{jk} + \tilde{\sigma}_{jk}\right)\partial_j\overline v_k \label{eq:mean-totE}
\end{multline}
Neglecting viscocity transport, we need to determine $\tilde{\sigma}_{ij}$, $\tilde{\omega}$, $F_{\rm c} = \overline{\rho} \overline{h'v_j'}$, and $\mathcal{F}_{\tilde{\omega}} = (1/2)\overline\rho\,\overline{v_j' v_k'v_k'}$ to close Eq. (\ref{eq:mean-mom2}) and (\ref{eq:mean-totE}). Since $\tilde{\omega}$  is half of the trace of $\tilde{\sigma}$, we derive the overall equation, and then derive $\tilde{\omega}$. First, we multiply (\ref{eq:fluc_mom}) with $v_i'$, then we take the some of this equation and the one with switched indices. Lastly, we take the average. The result after rearranging is:
\begin{multline}
    d_t \overline{v_i'v_j'} = -\overline{v_k'\partial_k(v_i'v_j')} - \left(\overline{v_i'v_k'}\,\partial_k \overline v_j+ \overline{v_j'v_k'}\,\partial_k \overline v_i\right) -\frac{1}{\overline \rho} \left(\overline{v_i'\partial_jp'}+\overline{v_j'\partial_i p'}\right)\\
    +\frac{1}{\overline{\rho}^2}\left(\overline{v_i'\rho'}\,\partial_j\overline p+\overline{v_j'\rho'}\partial_i\overline p\right) + \frac{1}{\overline \rho}\left(\overline {v_i'\partial_k\sigma_{kj}'}+ \overline{v_j'\partial_k\sigma'_{ki}}\right)\label{eq:Reynolds-full}.
\end{multline}
The first term of the right-hand side is the turbulent flux of the tensor component, the second term is the so-called shear production \citep{Pope-konyv}, which converts kinetic energy into turbulence. The third term can be split into two parts as $v_i'\partial_jp' = \partial_j(v_i'p')-p'\partial_jv_i'$. The first term describes the pressure transport, and the second term is the pressure-velocity rate-of-strain tensor. The role of this latter is to restore the isotropy of the turbulence. The last two terms are the bouyancy source term, which converts internal energy into turbulent energy, and the viscous dissipation, which again is composed of a transport term and a sink term, which generates heat ($v_i'\partial_k\sigma_{kj}' = \partial_k(v_i'\sigma_{kj}')-\sigma_{kj}'\partial_kv'_i$). Let the tensor, pressure, and viscosity transport term be note by $\mathcal{T}$, $\mathcal{T}^p$, $\mathcal{T}^\sigma$, the production term by $\mathcal{P}$, the pressure-velocity rate-of-strain tensor by $\Pi_{ij}$, the source by $\mathcal{S}$ and the dissipation by $\varepsilon$. Eq. (\ref{eq:Reynolds-full}) is then
\begin{multline}
    d_t \overline{v_i'v_j'} = -\frac{1}{\overline \rho} \partial_k\left(\mathcal{T}_{kij}+\mathcal{T}^p_{kij}-\mathcal T^{\sigma}_{kij}\right) + \mathcal{P}_{ij} +\Pi_{ij} +\mathcal{S}_{ij}-\varepsilon_{ij}.\label{eq:Reynolds-nice}
\end{multline}
Eq. (\ref{eq:Reynolds-nice}) shows, that the eddy viscosity hipothesis, that $\tilde{\sigma}_{ij}-(2/3)\overline\rho\,\tilde{\omega} \propto \partial_i\overline{v}_j$ is a very crude approximation. Therefore, \citet{Canuto1998} developed a model that takes into account these terms, including rotation, and calibrated it to observational and simulation results.

The half of the trace of Eq. (\ref{eq:Reynolds-nice}) gives the equation for the specific TKE , $\tilde{\omega}$, which is
\begin{multline}
    d_t\tilde{\omega} = - \frac{1}{\overline \rho}\partial_k \left(\frac{1}{2}\overline{\rho}\,\overline{v_k'v_j'v_j'} + \overline{v_j'p'} - \overline{v_j'\sigma_{kj}'} \right)\\ + \frac{\tilde{\sigma}_{kj}}{\overline{\rho}}\partial_k\overline{v}_j + \frac{\partial_j \overline p}{\overline\rho^2}\overline{v_j'\rho'} - \overline{\sigma_{kj}'\partial_kv_j'},\label{eq:omega1}
\end{multline}
where $\Pi_{jj} \approx 0$ was used\footnote{Strictly speaking, the trace of the pressure-rate-of-strain tensor is zero if $\partial_kv_k' =0$, hence in our case, a better description would be $\Pi_{jj} = \overline{v_j'p'}(\partial_j\ln\overline{\rho})$, which on the other hand, can be absorbed into the transport terms.}.

To close our system of equations, we need to define the remaining primed quantities. First, we model the enthalpy flux. Using the Taylor series
\begin{equation}
    h(\overline{s}+s',\overline{p}+p' )\approx h(\overline{s},\overline p) + \left.\frac{\partial h}{\partial s}\right\vert_p s' + \left. \frac{\partial h}{\partial p}\right\vert_s p' +\mathcal{O}(s'^2,p'^2), \label{eq:enthalpy-taylor}
\end{equation}
 and under the small Mach-number approximation \citep{Kuhfuss1986}, we have $h' \approx \overline{T}s'$. One can use this argument to produce the source term of Eq. (\ref{eq:omega1}) as $\overline{\rho'v_j'} \approx -(\overline{\delta}\overline{\rho}/\overline{c_p}) \overline{s'v_j'}$. In short, we reduce the two terms to a function of ${\rm \Pi}=\overline{v_j' s'}$. To produce an equation for $\rm \Pi$, one needs an equation for $s'$ as well. First, we divide Eq. (\ref{eq:heat}) by $T$ and take the average, thus:
 \begin{multline}
     d_t \overline{s} + \overline{v'_k\partial_ks'} = \frac{1}{\overline{\rho}\overline{T}}\left( - \partial_i\overline{F_i} + \overline{\sigma_{ij}}\partial_i\overline{v}_j+\overline{\sigma'_{ij}\partial_iv_j'} \right) \\-\frac{1}{\overline\rho \overline{T}^2}\left(-\overline{T'\partial_i F_i'} + \overline{T'(\sigma_ij\partial_iv_j)'} \right),\label{eq:mean-ent}
 \end{multline}
 where the second term of the right-hand side provides the coupling with the entropy fluctuations. We get the equation of the fluctuating entropy by taking the difference of Eq. (\ref{eq:heat}) and (\ref{eq:mean-ent}), thus
 \begin{equation}
     d_t s' + v_k'\partial_k\overline{s} + (v_k'\partial_ks')' = -\frac{1}{\overline{\rho}\overline{T}}\left[\partial_iF_i' - (\sigma_{ij}'\partial_i v'_j)'\right] + \text{higher order terms}. \label{eq:fluc-ent}.
 \end{equation}
 One produces the equation of $\rm \Pi$ by multiplying Eq. (\ref{eq:fluc_mom}) by $s'$ and Eq. (\ref{eq:fluc-ent}) by $v_j'$, and take the average of the sum. 
 This is
 \begin{multline}
     {\rm d}_t {\rm \Pi} = -\frac{1}{\overline{\rho}}\partial_k\left(\overline{\rho}\,\overline{v_k' v_j' s'}\right) - \frac{1}{\overline{\rho}}\tilde{\sigma}_{jk}\partial_k\overline s - \overline{v_k' s'}\partial_k \overline{v}_j \\- \overline{\left(\frac{1}{\rho}\partial_jp\right)'s'} + \frac{1}{\overline{\rho}}\overline{s'\partial_k\sigma_{kj}'}-\frac{1}{\overline{\rho}\overline T}\left[\overline{v_j'\partial_iF'_i}-\overline{v_j'(\sigma_{ik}'\partial_iv_k')'}\right]. \label{eq:pi_temp}
 \end{multline}
 Here, Eq. (\ref{eq:anel}) was used in the first term. Neglecting $p'$, the fourth term of the RHS reduces to $\overline{s'^2} \overline{\delta}/(\overline{\rho}\,\overline{c_p})\partial_j\overline{p}$, hence we produce our last equation to ${\rm \Phi}=\overline{s'^2}/2$ by multiplying Eq. (\ref{eq:fluc-ent}) by $s'$ and take the average, that is
 \begin{equation}
     {\rm d}_t {\rm \Phi} = - \frac{1}{2\overline{\rho}}\partial_k\left(\overline{\rho}\,\overline{v_k's'^2}\right) - \overline{s'v_k'}\partial_k\overline{s}-\frac{1}{\overline\rho\,\overline T}\left[\overline{s'\partial_iF_i'} - \overline{s'(\sigma_{ij}'\partial_iv_j')'}\right]. \label{eq:phi_temp}
 \end{equation}

The remaining terms need to be closed algebraically because we want to produce a three-equation model. 

Firstly, we consider the average and fluctuating radiative flux. The flux can be written as $F_i = K_r(T,\kappa) \partial_iT$, then its average and first-order fluctuation is:
\begin{align}
    \overline{F_i} &= \overline{K_r}\,\partial_i\overline{T} + \overline{K_r'\partial_iT'}, \\
    F_i'& =K_r' \partial_i \overline{T} + \overline{K_r} \partial_i T',
\end{align}
where, introducing $\kappa_T = (\partial\ln\kappa/\partial\ln T)_p$, we have

\begin{equation}
    K_r' \approx \frac{\overline{K_r}}{\overline{c_p}}\left(3-\kappa_T\right)s'.
\end{equation}

To describe the radiation dissipation effects, we have to model $\partial_iT' \approx \overline{T}/\overline{c_p} \partial_i s'$. To do this, following \citet{Kuhfuss1987}, we introduce the radiation dissipation length $\Lambda_r$, hence $\partial_i s' \approx s'/\Lambda_r$. This means, that the radiation fluctuation terms are modeled as:
\begin{displaymath}
   \frac{1}{\overline \rho \overline T} \overline{s' \partial_i\overline{K_r}\partial_i T'} \approx \frac{2\Phi}{\tau_r},\qquad \frac{1}{\overline \rho \overline T}\overline{v_j'\partial_i\overline{K_r}\partial_iT'}  \approx \frac{\rm \Pi}{\tau_r},
\end{displaymath}
\begin{displaymath}
 \frac{1}{\overline \rho }\partial_i\overline{K_r'\partial_i T'} \approx \frac{2\Phi\overline{T}}{\overline{c_p}\tau_{\kappa}},\qquad  \frac{1}{\overline \rho \overline T}\overline{s'\partial_i K_r'\partial_i\overline T} \approx \frac{2\rm \Phi}{\tau_\kappa}, \qquad  \frac{1}{\overline \rho \overline T}\overline{v_j' \partial_i K_r' \partial_i \overline{T}}\approx \frac{\rm \Pi}{\tau_\kappa},
\end{displaymath}
where
\begin{displaymath}
    \tau_r = \frac{\Lambda_r^2\overline{\rho}\, \overline{c_p}}{\overline{K_r}\,},\quad \tau_\kappa=\frac{\tau_r}{3-\kappa_T}.
\end{displaymath}
Here we note that although $\tau_\kappa$ has the dimension of time, it is not a simple dissipation timescale, as when $\kappa_T < -3$, it provides an additional source of entropy-fluctuation. 

The next term we model is the dissipation of the TKE. We use the local $\tau_{\rm conv}=\Lambda\tilde{\omega}^{-1/2}$ approximation leading to
\begin{displaymath}
    \overline{\sigma_{ij}'\partial_i v_j'} \approx \frac{\tilde{\omega}}{\tau_{\rm conv}} =\alpha_d \frac{\tilde{\omega}^{3/2}}{\Lambda}.
\end{displaymath}

We consider the dynamic dissipation of $\Pi$ and $\rm \Phi$ to happen on similar scale then on $\tilde{\omega}$ thus
\begin{align*}
    \frac{1}{\overline\rho}\overline{s'\partial_k\sigma_{kj}'}-\frac{1}{\overline\rho \overline{T}}\overline{v_j'(\sigma_{ik}'\partial_iv_k')'} &\approx - \alpha_{d,\Pi}\frac{\Pi \tilde{\omega}^{1/2}}{\Lambda}\\
    -\frac{1}{\overline \rho \overline T }\overline{s'(\sigma_{ij}'\partial_iv_j')'} &\approx -\alpha_{\rm d,\Phi}\frac{\Phi \omega^{1/2}}{\Lambda}
\end{align*}

Now, we consider the property of averaging that only keeps radial terms of the vectors. This means that $\overline{v_j s'} =\overline{v_3 s'}$, and $\partial_j \overline{p} =\partial_3\overline{p} \delta_{j3}$.

Furthermore, we define anisotropy as $\xi = \overline{v_3'^2}/(2\tilde{\omega})$, and assume horizontal isotropy thus, $\overline{v_1'^2} = \overline{v_2'^2} = \tilde{\omega}(1-\xi)$. Using this, we split the Reynolds tensor to the isotropic turbulent pressure, and anisotropic eddy viscous pressure \citep{GW1992}: $-\tilde{\sigma}_{ij} = (2/3)\tilde{\omega}\,\overline{\rho}\delta_{ij}+P^\nu_{ij}$, so only $P_{ij}^\nu$ is need to be modeled, which equivalent with $\tilde{\omega}b_{ij}$ in \citet{Canuto1998} notation. Contributions of the eddy viscous pressure in Eq. (\ref{eq:mean-mom2}) and Eq. (\ref{eq:mean-totE}) are called turbulent momentum transfer rate ($U_\nu$) and turbulent viscous energy transfer rate ($E_\nu$), respectively \citep{Wuchterl1998}. Their definitions ar
\begin{align}
    U_\nu &\coloneqq \frac{1}{\overline\rho} \partial_k P_{kj}^\nu, & E_\nu &\coloneqq \frac{1}{\overline{\rho}} P_{ij}^\nu\partial_i \overline{v}_j. &
\end{align}

Lastly, we describe the TOMs. In Eq. (\ref{eq:omega1}) we absorb pressure, viscosity transport, and the TOM into a single diffusion flux term using the diffusion approximation, we introduce the turbulent dynamic viscosity as $\mu_{\rm t} \coloneqq \overline{\rho}\Lambda \tilde{\omega}^{1/2}$, thus 
\begin{equation}
    \mathcal{F}_{\tilde{\omega}} \coloneqq\frac{1}{2} \overline \rho \overline{v_k' v_j'v_j'} + \overline{v_k'p'} + \overline{v_j'\sigma_{kj}'} \approx -\alpha_t \xi \mu_{\rm t}\partial_r\,\tilde{\omega}. 
\end{equation}
The remaining TOMs in eqs. (\ref{eq:pi_temp})-(\ref{eq:phi_temp}) become
\begin{align}
    \mathcal F_{\rm \Pi} &\coloneqq \overline{\rho}\,\overline{v_k' v_j' s'} \approx -\alpha_{\rm \Pi}\xi \mu_{\rm t}\partial_r{\rm \Pi}\\
    \mathcal F_{\rm \Phi} &\coloneqq \frac{\overline{\rho}}{2}\,\overline{v'_ks'^2} \approx -\alpha_\Phi \xi \mu_t \partial_r\Phi
\end{align}

The model equations presented in Sect. \ref{sec:model}. can be derived by using spherical symmetry and substituting all previously described terms into Eqs. (\ref{eq:mean_kont}), (\ref{eq:mean-mom2}), (\ref{eq:mean-totE}), (\ref{eq:omega1}),  (\ref{eq:pi_temp}), and (\ref{eq:phi_temp}).

\section{Preliminaries for the anisotropy expression}
\label{ap:anisotropy}
From \citet{Canuto1998b} (25a) and (27a), one can derive the evolution equation of $\overline{v_3'^2}=-\overline{\rho}^{-1}\tilde\sigma_{33}$ using the same steps as \citet{Kupka2022}. To derive the shear terms, one has to use the definition of the mean velocity, that is $\overline{v}_i = \partial_{i3}\overline{v}(r)$. Therefore
$$\partial_i\overline{v}_j = \left(\begin{matrix}
    \overline{v}/r &0& 0\\
    0& \overline{v}/r & 0 \\
    0&0&\partial_r v_r
\end{matrix} \right)$$ and the flux term is $\lambda_ih_j = \delta_{i3}\delta_{ij}S$. Furthermore, we assume isotropy in the horizontal direction, thus $\overline{v_1'^2}+\overline{v_2'^2} = 2\tilde{\omega}-\overline{v_3'^2}$.

\section{Derivation of the value of $\alpha_c$}
\label{ap:convection}

The modeling of the convective flux is based on the Taylor expansion of the enthalpy as a function of temperature,
which is equivalent to eq (\ref{eq:enthalpy-taylor}) with using: $T'\approx (\overline{T}/\overline{c_p}) s'$. 
On the other hand, the transport of ions means that the mean molecular weight also changes, that is,
\begin{multline}
h(\overline T + T', \overline p +p', \overline{\mu}+\mu' ) = h(\overline{T},\overline{p})\\+ \left.\frac{\partial h}{\partial T} \right\vert_p T' + \left. \frac{\partial h}{\partial p}\right\vert_T p' + \left.\frac{\partial h}{\partial \mu}\right\vert_{T_,p}\mu'+ \mathcal{O}(T'^2 , p'^2,\mu'^2)
\end{multline}
The absorption of ionization effects into $c_p=(\partial h/\partial T)_p$ and $(\partial h/\partial p )_T$ is a usual process \citep{Kippenhahn-book}, but in this case the specific heat is
\begin{equation}
    c_p \equiv \left. \frac{\partial h}{\partial T} \right\vert_p = h_T + h_\mu \mu_T,
\end{equation}
where $h_T=({\partial h}/{\partial T} )_{p,\mu}$, $h_\mu=({\partial h}/{\partial \mu} )_{p,t}$, $\mu_T=({\partial \mu}/{\partial T} )_{p}$. If we use Reynolds-averaging, we get
\begin{align}
    \overline{c_p} &= \overline{h_T} + \overline{h_\mu}\overline{\mu_T} + \overline{h_\mu'\mu_T'} = c_p(\overline{T},\overline{p})+\overline{h_\mu' \mu_T'},\\
    c_p' &= h_T' + \overline{h_\mu}\mu_T' + h_\mu' \overline{\mu_T} + (h_\mu' \mu_T')'. \label{eq:cpfluc}
\end{align}
Now, we assume that all fluctuations are due to temperature fluctuation in the first order. Thus $c_p' = c_{pT}T'$, $h_T'=h_{TT} T'$ $h_{\mu}'=h_{\mu T} T'$, where $T$ in the subscript means derivative by $T$, for example, $h_{TT} = (\partial^2 h/\partial T^2)_p$, and all second derivatives are considered to be functions of average values only\footnote{We note that  we do not use dimensionless derivatives in this derivation, in contrast with Sect. \ref{sec:model}.}. Substituting these into (\ref{eq:cpfluc}) and dividing by $T'$ we get
\begin{equation}
    c_{pT} = h_{TT} + \overline{h_\mu}\mu_{TT} + h_{\mu T} \overline{\mu_T} + h_{\mu T} \mu_{T T} T' - h_{\mu T} \mu_{T T} \overline{T'^2}/T'
\end{equation}

Let $c_{p0} = c_p (\overline T, \overline p) = \overline{h_T} + \overline{h_\mu}\overline{\mu_T}$ and $c_{pT0}= \overline{h_{TT}} + \overline{h_{\mu T} }\overline{\mu_T} + \overline{h_\mu} \overline{\mu_{TT}}$. These are the specific values that can be exactly determined from the mean values, and as we can see, $c_p \neq c_{p0}$. Now substitute these back into the definition of $h'$, and we get
\begin{equation}
    h' \approx c_{p0} T' + h_{\mu T}\mu_{TT} \overline{T'^2} T' + c_{pT0} T'^2 + h_{\mu T}\mu_{TT}  T'^3 -h_{\mu T}\mu_{TT}\overline{T'^2}T', 
\end{equation}
therefore setting $\alpha_c=1$ is adequate to follow ionization changes in first-order.

\end{appendix}
\end{document}